\documentclass[journal]{IEEEtran}
\usepackage{graphicx,amsmath,url}
\usepackage{microtype,lmodern}
\usepackage{mathtools,amssymb,gensymb}
\usepackage{tgtermes} 
\usepackage[T1]{fontenc}
\usepackage[utf8]{inputenc}
\usepackage{color,soul}
\usepackage[colorinlistoftodos]{todonotes}
\usepackage{acronym}
\usepackage{array}
\usepackage{booktabs}
\usepackage{subcaption}
\usepackage[compress]{cite}

\usepackage[normalem]{ulem}
\usepackage[switch]{lineno}
\usepackage{listings}
 
\usepackage{longtable}
\usepackage[colorlinks,breaklinks]{hyperref} 
\usepackage{acronym}
\usepackage{graphicx}

\hypersetup{hypertexnames=true,
  linkcolor=blue,anchorcolor=black,citecolor=blue,urlcolor=blue
}
\graphicspath{{imglocal/}{img/}}
\acrodef{ADC}[ADC]{Analog to Digital Converter}
\acrodef{ADEX}[AdExp-I\&F]{Adaptive-Exponential Integrate and Fire}
\acrodef{AER}[AER]{Address-Event Representation}
\acrodef{AEX}[AEX]{AER EXtension board}
\acrodef{AE}[AE]{Address-Event}
\acrodef{AFM}[AFM]{Atomic Force Microscope}
\acrodef{AMDA}[AMDA]{AER Motherboard with D/A converters}
\acrodef{ANN}[ANN]{Attractor Neural Network}
\acrodef{API}[API]{Application Programming Interface}
\acrodef{ARM}[ARM]{Advanced RISC Machine}
\acrodef{ASIC}[ASIC]{Application Specific Integrated Circuit}
\acrodef{BCM}[BMC]{Bienenstock-Cooper-Munro}
\acrodef{BD}[BD]{Bundled Data}
\acrodef{BEOL}[BEOL]{Back-end of Line}
\acrodef{BG}[BG]{Bias Generator}
\acrodef{BMI}[BMI]{Brain-Machince Interface}
\acrodef{CAD}[CAD]{Computer Aided Design}
\acrodef{CAM}[CAM]{Content Addressable Memory}
\acrodef{CAVIAR}[CAVIAR]{Convolution AER Vision Architecture for Real-Time}
\acrodef{CCN}[CCN]{Cooperative and Competitive Network}
\acrodef{CMOL}[CMOL]{``Hybrid CMOS nanoelectronic circuits''}
\acrodef{CMIM}[CMIM]{Metal-insulator-metal Capacitor}
\acrodef{CMOS}[CMOS]{Complementary Metal-Oxide-Semiconductor}
\acrodef{COTS}[COTS]{Commercial Off-The-Shelf}
\acrodef{CPG}[CPG]{Central Pattern Generator}
\acrodef{CPLD}[CPLD]{Complex Programmable Logic Device}
\acrodef{CPU}[CPU]{Central Processing Unit}
\acrodef{CV}[CV]{Coefficient of Variation}
\acrodef{DAC}[DAC]{Digital to Analog Converter}
\acrodef{DAS}[DAS]{Dynamic Auditory Sensor}
\acrodef{DAVIS}[DAVIS]{Dynamic and Active Pixel Vision Sensor}
\acrodef{DBN}[DBN]{Deep Belief Network}
\acrodef{DFA}[DFA]{Deterministic Finite Automaton}
\acrodef{DMA}[DMA]{Direct Memory Access}
\acrodef{DNF}[DNF]{Dynamic Neural Field}
\acrodef{DNN}[DNN]{Deep Neural Network}
\acrodef{DOF}[DOF]{Degrees of Freedom}
\acrodef{DPE}[DPE]{Dynamic Parameter Estimation}
\acrodef{DPI}[DPI]{Differential Pair Integrator}
\acrodef{DRAM}[DRAM]{Dynamic Random Access Memory}
\acrodef{DR}[DR]{Dual Rail}
\acrodef{DSP}[DSP]{Digital Signal Processor}
\acrodef{DVS}[DVS]{Dynamic Vision Sensor}
\acrodef{EBL}[EBL]{Electron Beam Lithography}
\acrodef{EDVAC}[EDVAC]{Electronic Discrete Variable Automatic Computer}
\acrodef{EIN}[EIN]{Excitatory-Inhibitory Network}
\acrodef{EM}[EM]{Expectation Maximization}
\acrodef{EPSC}[EPSC]{Excitatory Post-Synaptic Current}
\acrodef{EPSP}[EPSP]{Excitatory Post-Synaptic Potential}
\acrodef{FD-SOI}[FD-SOI]{Fully-Depleted Silicon on Insulator}
\acrodef{FET}[FET]{Field-Effect Transistor}
\acrodef{FFT}[FFT]{Fast Fourier Transform}
\acrodef{FI}[F-I]{Frequency-Current}
\acrodef{FPGA}[FPGA]{Field Programmable Gate Array}
\acrodef{FSA}[FSA]{Finite State Automaton}
\acrodef{FSM}[FSM]{Finite State Machine}
\acrodef{GOPS}[GOPS]{Giga-Operations per Second}
\acrodef{GPU}[GPU]{Graphical Processing Unit}
\acrodef{GUI}[GUI]{Graphical User Interface}
\acrodef{HAL}[HAL]{Hardware Abstraction Layer}
\acrodef{HH}[H\&H]{Hodgkin \& Huxley}
\acrodef{HMM}[HMM]{Hidden Markov Model}
\acrodef{HRS}[HRS]{High-Resistive State}
\acrodef{HR}[HR]{Human Readable}
\acrodef{HW}[HW]{Hardware}
\acrodef{ICT}[ICT]{Information and Communication Technology}
\acrodef{IC}[IC]{Integrated Circuit}
\acrodef{IF2DWTA}[IF2DWTA]{Integrate \& Fire 2--Dimensional WTA}
\acrodef{IFSLWTA}[IFSLWTA]{Integrate \& Fire Stop Learning WTA}
\acrodef{IF}[I\&F]{Integrate-and-Fire}
\acrodef{IMU}[IMU]{Inertial Measurement Unit}
\acrodef{INCF}[INCF]{International Neuroinformatics Coordinating Facility}
\acrodef{INI}[INI]{Institute of Neuroinformatics}
\acrodef{IO}[I/O]{Input/Output}
\acrodef{IPSC}[IPSC]{Inhibitory Post-Synaptic Current}
\acrodef{IPSP}[IPSP]{Inhibitory Post-Synaptic Potential}
\acrodef{IP}[IP]{Intellectual Property}
\acrodef{ISI}[ISI]{Inter-Spike Interval}
\acrodef{JFLAP}[JFLAP]{Java - Formal Languages and Automata Package}
\acrodef{LFP}[LFP]{Local Field Potential}
\acrodef{LNA}[LNA]{Low-Noise Amplifier}
\acrodef{LPF}[LPF]{Low-Pass Filter}
\acrodef{LRS}[LRS]{Low-Resistive State}
\acrodef{LSM}[LSM]{Liquid State Machine}
\acrodef{LTD}[LTD]{Long Term Depression}
\acrodef{LTI}[LTI]{Linear Time-Invariant}
\acrodef{LTP}[LTP]{Long Term Potentiation}
\acrodef{LTU}[LTU]{Linear Threshold Unit}
\acrodef{LUT}[LUT]{Look-Up Table}
\acrodef{MCMC}[MCMC]{Markov-Chain Monte Carlo}
\acrodef{MEMS}[MEMS]{Micro Electro Mechanical System}
\acrodef{MIM}[MIM]{Metal Insulator Metal}
\acrodef{MOS}[MOS]{Metal Oxide Semiconductor}
\acrodef{MOSCAP}[MOSCAP]{Metal Oxide Semiconductor Capacitor}
\acrodef{MOSFET}[MOSFET]{Metal Oxide Semiconductor Field-Effect Transistor}
\acrodef{MRI}[MRI]{Magnetic Resonance Imaging}
\acrodef{NDFSM}[NDFSM]{Non-deterministic Finite State Machine} 
\acrodef{ND}[ND]{Noise-Driven}
\acrodef{NEF}[NEF]{Neural Engineering Framework}
\acrodef{NHML}[NHML]{Neuromorphic Hardware Mark-up Language}
\acrodef{NIL}[NIL]{Nano-Imprint Lithography}
\acrodef{NMDA}[NMDA]{N-Methyl-D-Aspartate}
\acrodef{NME}[NE]{Neuromorphic Engineering}
\acrodef{OTA}[OTA]{Operational Transconductance Amplifier}
\acrodef{PCB}[PCB]{Printed Circuit Board}
\acrodef{PCHB}[PCHB]{Pre-Charge Half-Buffer}
\acrodef{PSC}[PSC]{Post-Synaptic Current}
\acrodef{PFM}[PFM]{Pulse Frequency Modulation}
\acrodef{PSTH}[PSTH]{Peri-Stimulus Time Histogram}
\acrodef{QDI}[QDI]{Quasi-Delay Insensitive}
\acrodef{RAM}[RAM]{Random Access Memory}
\acrodefplural{RAM}[RAMs]{Random Access Memories}
\acrodef{RMSE}[RMSE]{Root Mean Squared-Error}
\acrodef{RMS}[RMS]{Root Mean Squared}
\acrodef{RNN}[RNN]{Recurrent Neural Network}
\acrodef{ROLLS}[ROLLS]{Reconfigurable On-Line Learning Spiking}
\acrodef{RRAM}[RRAM]{Resistive Random Access Memory}
\acrodefplural{RRAM}[R-RAMs]{Resistive Random Access Memories}
\acrodef{SAC}[SAC]{Selective Attention Chip}
\acrodef{SCX}[SCX]{Silicon CorteX}
\acrodef{SD}[SD]{Signal-Driven}
\acrodef{SEM}[SEM]{Spike-based Expectation Maximization}
\acrodef{SLAM}[SLAM]{Simultaneous Localization and Mapping}
\acrodef{SOI}[SOI]{Silicon on Insulator}
\acrodef{SOC}[SOC]{System-On-Chip}
\acrodef{SRAM}[SRAM]{Static Random Access Memory}
\acrodef{STDP}[STDP]{Spike-Timing Dependent Plasticity}
\acrodef{STD}[STD]{Short-Term Depression}
\acrodef{STP}[STP]{Short-Term Plasticity}
\acrodef{STT}[STT]{Spin-Transfer Torque}
\acrodef{STT-MRAM}[STT-MRAM]{Spin-Transfer Torque Magnetic Random Access Memory}
\acrodefplural{STT-MRAM}[STT-MRAMs]{Spin-Transfer Torque Magnetic Random Access Memories}
\acrodef{SW}[SW]{Software}
\acrodef{TFT}[TFT]{Thin Film Transistor}
\acrodef{USB}[USB]{Universal Serial Bus}
\acrodef{VHDL}[VHDL]{VHSIC Hardware Description Language}
\acrodef{VLSI}[VLSI]{Very Large Scale Integration}
\acrodef{VOR}[VOR]{Vestibulo-Ocular Reflex}
\acrodef{WTA}[WTA]{Winner-Take-All}
\acrodef{XML}[XML]{eXtensible Mark-up Language}
\acrodef{divmod3}[DIVMOD3]{divisibility of a number by 3}
\acrodef{hWTA}[hWTA]{Hard Winner-Take-All}
\acrodef{sWTA}[sWTA]{soft Winner-Take-All}

\begin{document}
%
\title{Scaling mixed-signal neuromorphic processors to 28\,nm FD-SOI technologies}

\author{\IEEEauthorblockN{Ning Qiao}
\IEEEauthorblockA{Institute of Neuroinformatics\\
University of Zurich and ETH Zurich\\
Zurich, Switzerland\\
Email: qiaoning@ini.uzh.ch}
\and
\IEEEauthorblockN{Giacomo Indiveri}
\IEEEauthorblockA{Institute of Neuroinformatics\\
University of Zurich and ETH Zurich\\
Zurich, Switzerland\\
Email: giacomo@ini.uzh.ch}}


%


\maketitle

\begin{abstract}
As processes continue to scale aggressively, the design of deep sub-micron, mixed-signal design is becoming more and more challenging. In this paper we present an analysis of scaling multi-core mixed-signal neuromorphic processors to advanced 28\,nm \acs{FD-SOI} nodes. We address analog design issues which arise from the use of advanced process, including the problem of large leakage currents and device mismatch, and asynchronous digital design issues. We present the outcome of Monte Carlo Analysis and circuit simulations of neuromorphic subthreshold analog/digital neuron circuits which reproduce biologically plausible responses. We describe the \acs{AER} used to implement \acs{PCHB} based asynchronous \acs{QDI} routing processes in multi-core neuromorphic architectures and validate their operation via circuit  simulation results. Finally we describe the implementation of custom 28\,nm \acs{CAM} based memory resources utilized in these multi-core neuromorphic processor and discuss the possibility of increasing density by using advanced \acs{RRAM} devices integrated in the  28\,nm \ac{FD-SOI} process.   
\end{abstract}


%
\IEEEpeerreviewmaketitle

\acresetall

\section{Introduction}

Neural networks and deep learning models have recently become the state of the art architectures for a wide range of applications that include data mining, signal processing, and pattern recognition~\cite{LeCun_etal15}. 
However, most of these architectures are modeled as algorithms executed on power-hungry central processing units (CPUs) or graphical processing units (GPUs), often integrated in large server farms of classical von Neumann computing systems. 
Neuromorphic processors represent a promising alternative brain-inspired technology that, thanks to their massively parallel computational substrate, are ideally suited for implementing such algorithms~\cite{Merolla_etal14a,Benjamin_etal14,Furber_etal14,Chicca_etal14,Qiao_etal15}. These hardware devices promise to reduce power consumption by several orders of magnitude and have the potential to solve the von Neumann memory bottleneck problem thanks to their co-localized memory and computing features~\cite{Indiveri_Liu15}. 

While neuromorphic processors made using purely digital circuits are already being implemented in advanced scaled processes~\cite{Merolla_etal14a,Furber_etal14}, devices designed using mixed analog and digital circuits have been implemented typically using older, less aggressive CMOS processes, such as 180\,nm~\cite{Qiao_etal15,Benjamin_etal14}.

In this paper we show how subthreshold analog designs can be scaled effectively down to 28\,nm processes using \ac{FD-SOI} technology. We use as a reference design the analog silicon neuron described in~\cite{Qiao_etal15} and analyze its performance and characteristics in the 28\,nm \ac{FD-SOI} process, addressing issues related to channel and gate leakage currents, as well as device mismatch. Moreover, we  describe the set of asynchronous digital circuits that can be used to interconnect multiple instances of these neurons among each others in single-chip multi-core architectures, and provide estimates for  their size, bandwidth and power consumption in these scaled processes.




\section{Key sub-circuits in 28\,nm FD-SOI processes}

The transistors operated in the subthreshold regime used in mixed signal analog/digital neuromorphic architectures implemented with 180\,nm or larger feature size CMOS processes typically use currents ranging from tens of nA to currents as low as a few pA~\cite{Liu_etal02a}. Minimum-size transistors in advanced processes have considerably larger leakage currents.
To maintain these levels of currents,  we performed circuit simulations of single transistors and determined their proper geometrical size. Based on these results, we optimized the design of the analog silicon neuron circuit shown in Fig.~\ref{fig:soma}, and performed Monte Carlo simulations to validate its performance in the 28\,nm \ac{FD-SOI} process.  
\begin{figure*}
  \begin{subfigure}{0.45\textwidth}
    \centering
    \includegraphics[width=0.8\textwidth]{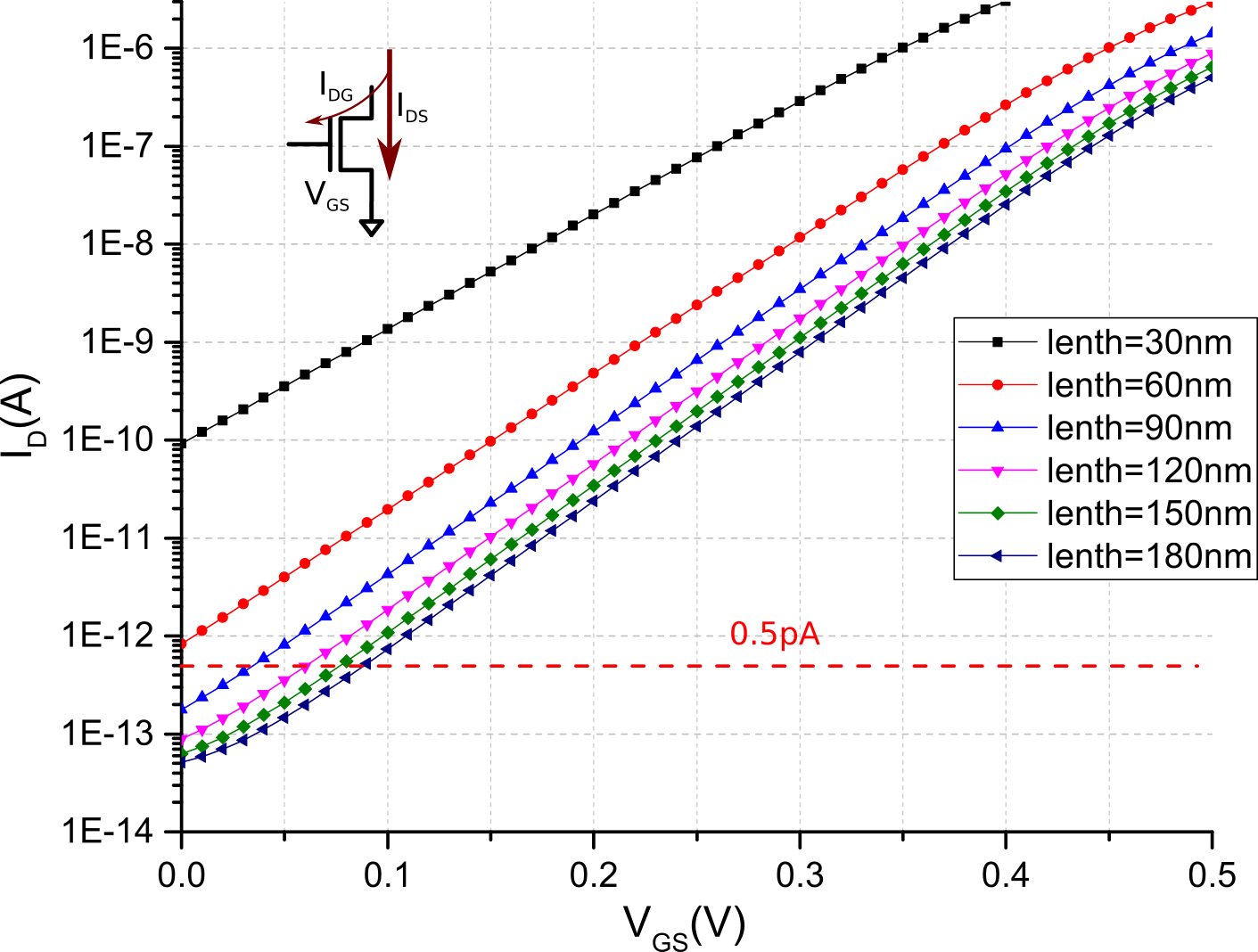}
    \subcaption{} 
    \label{fig:NMOS}
  \end{subfigure}
  \begin{subfigure}{0.45\textwidth}
    \centering
    \includegraphics[width=0.8\textwidth]{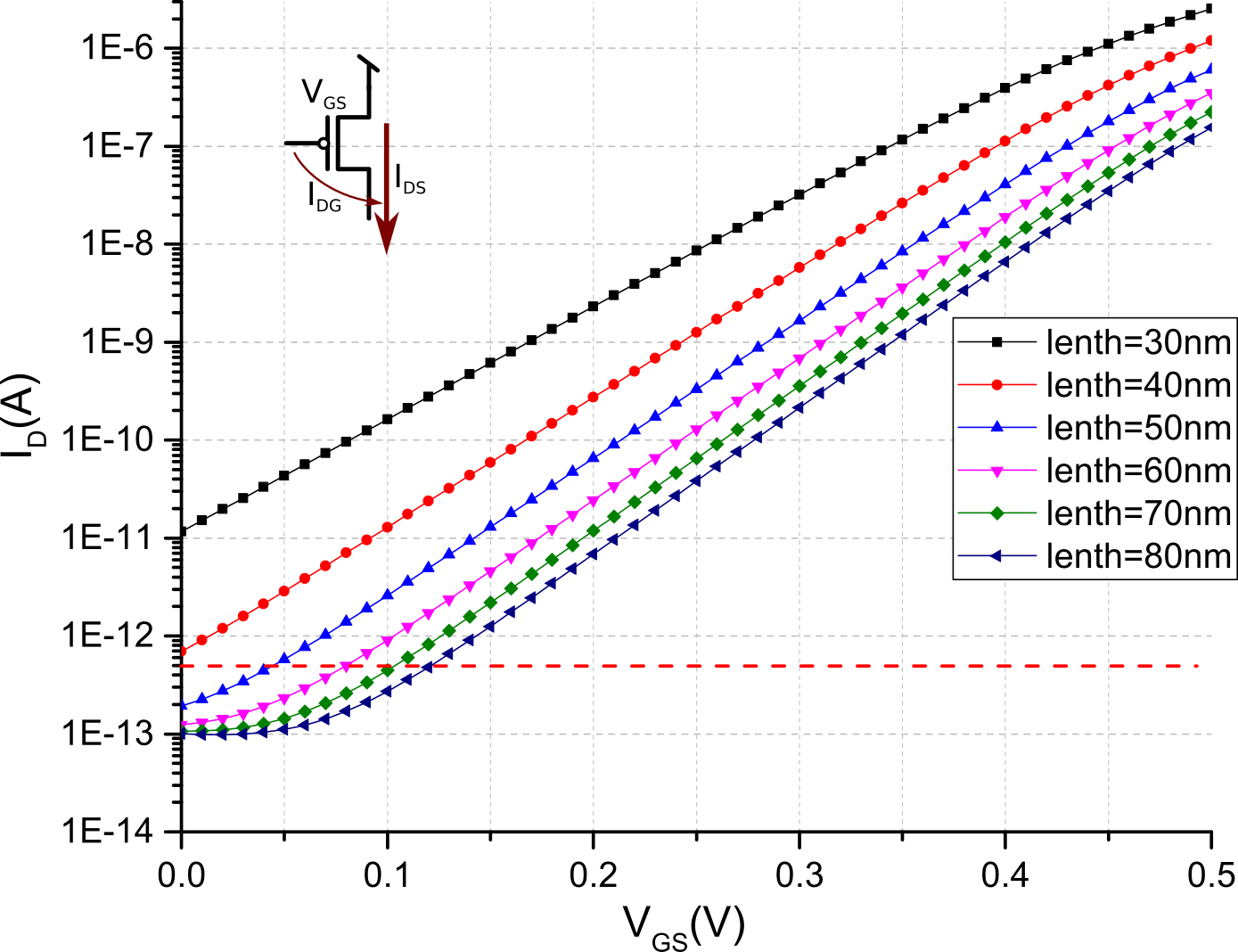}
    \subcaption{} 
    \label{fig:PMOS}
  \end{subfigure}
  \caption{Simulated channel current $I_{D}$ of versus  $V_{GS}$ in subthreshold region for different channel length with $V_{DS} = 0.5V$ and $W = 200\,nm$, for NMOS (a) and PMOS (b) transistors.}
  \label{fig:ids}
\end{figure*}

To build multi-neuron/multi-core architectures using these analog neuron circuits, we make use of asynchronous \ac{AER} digital circuits. These circuits assign a tag to the neuron that spikes and route its address to one or more destinations. Routing tables and tag memories are distributed within and across the neuron arrays, and can be programmed using the same \ac{AER} communication protocol. Different neural network configurations, such as convolutional networks, deep multi-layer networks, or recurrent reservoir networks, can be deployed, depending on how these memory structures are programmed.
To evaluate performance of typical asynchronous \ac{AER} circuits in the 28\,nm \ac{FD-SOI} process considered, we designed and simulated a 10-bit pipeline buffer process stage. 
The type of digital memories we considered for storing the tags used to route spikes from source neurons to destination synapses are \acp{CAM}. As these are integral part of the synapse modules, and as neuromorphic processors typically consist of large arrays of synaptic elements, these memory circuits are the ones that would occupy the main area of the chip. We discuss possible scaling strategies for implementing \acp{CAM} in order to minimize the area of the die size, in the 28\,nm \ac{FD-SOI} process, given a specific network size. 

\subsection{The analog subthreshold Integrate\&Fire neuron}

\begin{figure}
  \centering
  \includegraphics[width=0.44\textwidth]{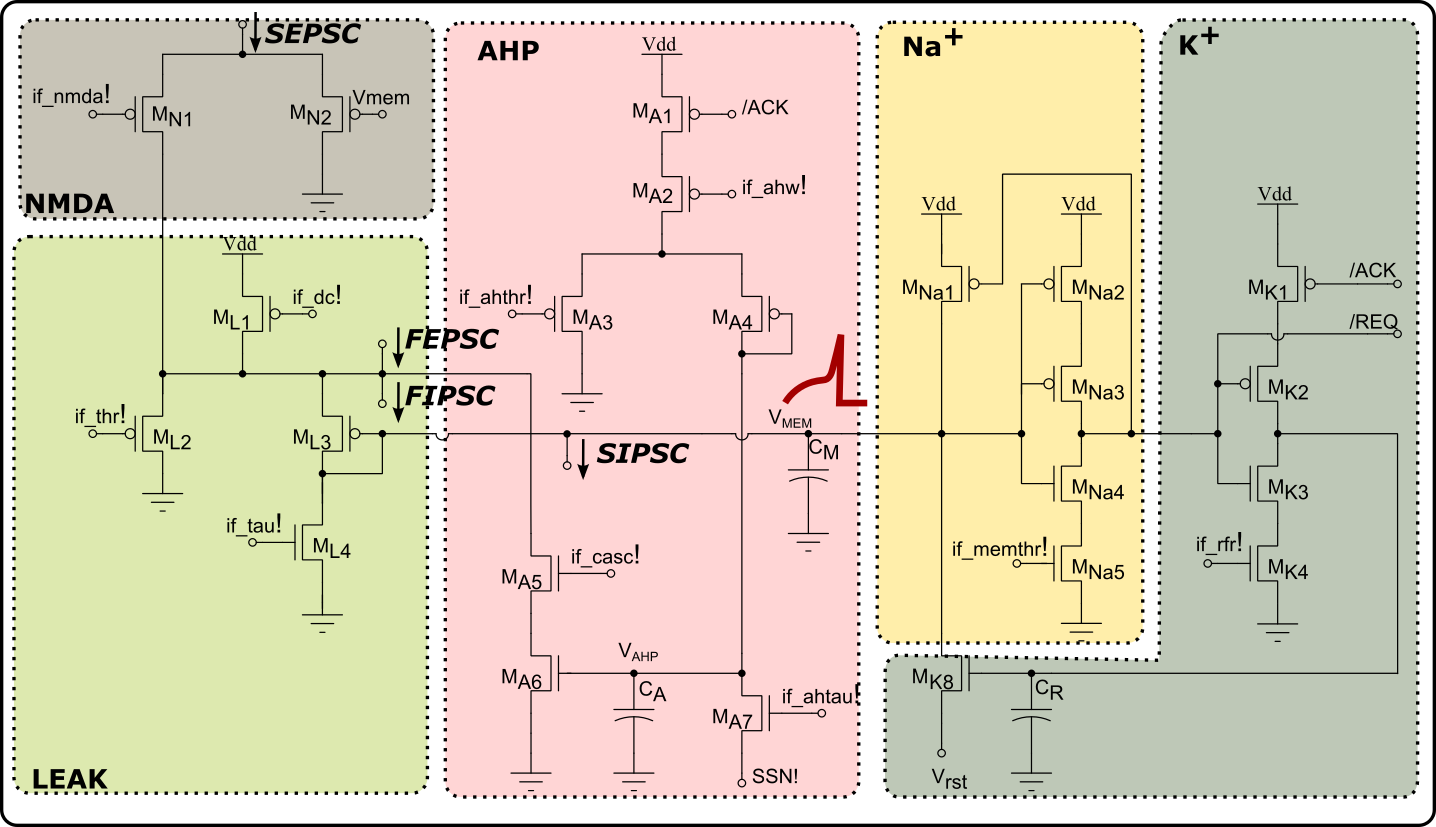}
  \caption{Simplified schematic diagram of an analog I\&F neuron.}
  \label{fig:soma}
\end{figure}

To develop mixed signal analog/digital neuromorphic processors that can be used in closed-loop application scenarios (e.g. ranging from self-driving cars to biomedical micro-devices measuring metabolites in the blood stream and deciding on what actions to take), it is necessary to endow them with computational elements that have time constants that are well-matched to the signals they are meant to process~\cite{Chicca_etal14}. For natural signals (speech, gestures, etc.) these signals typically involve time constants of the order of milliseconds. It has been shown~\cite{Qiao_etal15} that
to achieve these time constants with the silicon neuron of Fig.~\ref{fig:soma}, given a membrane capacitance of 1\,pF, it is necessary to use currents of the order of a few pA.
 
Figure~\ref{fig:ids} shows simulated current $I_{D}$ of PMOS/NMOS versus $|V_{GS}|$ in subthreshold region for different channel lengths, with a fixed $|V_{DS}| = 0.5V$ and $W = 200\,nm$. 

A simplified subthreshold analog neuron circuit compatible with the 28\,nm \ac{FD-SOI} process is shown in Fig.~\ref{fig:soma}. Input currents $I_{syn}$ are injected into the neuron membrane capacitance $C_{M}$, in parallel with a programmable constant DC current. The \emph{NMDA} block models the voltage-gating mechanisms of NMDA synapses. The \emph{LEAK} block models the neuron's leak conductance. The \emph{AHP} block models the generation of the after hyper-polarizing current in real neurons, responsible for their spike-frequency adaptation behavior. The Na and K block model the effect of Sodium and Potassium channels, responsible for generating action-potentials (spikes) in real neurons. The $REQ$ and $ACK$ signals represent the digital voltages used to communicate Address-Events to the output \ac{AER} circuits. All signals ending with "!" represent global variables (shared parameters) used to set the neuron firing properties. The $I_{mem}$ and $I_{ah}$ currents represent the fast and slow variables in the AdExp model, respectively. As shown in Fig.~\ref{fig:soma}, the Fast Excitatory Post-Synaptic Current (FEPSC), Slow Excitatory Post-Synaptic Current (SEPSC), Fast Inhibitory Post-Synaptic Current (FIPSC) and Slow Inhibitory Post-Synaptic Current (SIPSC), with independently fine-tunable time constant parameters, feed into different branches of the neuron circuit.

\begin{figure}
  \centering
  \includegraphics[width=0.39\textwidth]{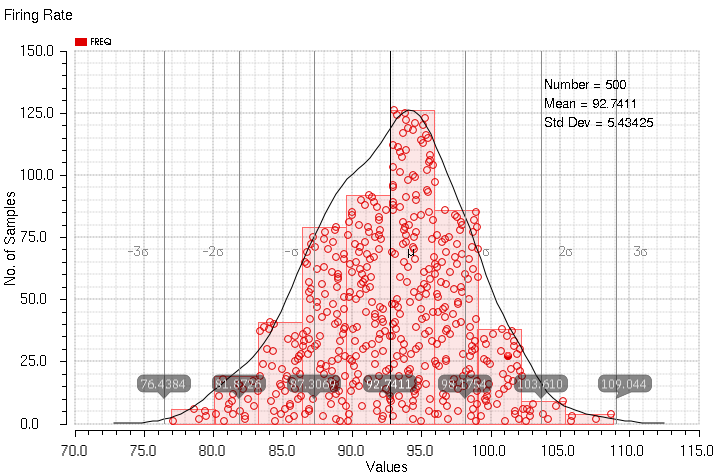}
  \caption{Monte Carlo analysis results for 500 runs with circuit parameters set in a way to obtain the neuron's mean firing rate centered at 92.74Hz and its standard deviation at 5.43, with relative error of firing rate ($Std\_Dev/Mean$) equal to 5.86\%}
  \label{fig:mismatch}
\end{figure}

To minimize transistor mismatch effects, we identified the ones that are required to operate with small currents and assigned them large length values (e.g., $L_{P} = 100\,nm$, $L_{N} = 200\,nm$). Even larger transistor sizes (e.g., $500\,nm/500\,nm$) were assigned key transistors relevant for mismatch (e.g., $M_{Na5}$ and $M_{K4}$) .
We performed Monte Carlo analysis with 500 runs for this neuron circuit, with DC  current injected through $M_{L1}$, and with bias voltages set to obtain a firing rate of approximately 100\,Hz. 
As shown in Fig.~\ref{fig:mismatch}, for a mean firing rate of 92.7Hz, the standard deviation is 5.43 and error ($Std\_Dev/Mean$) is 5.86\%.  

\begin{figure}
  \centering
  \includegraphics[width=0.43\textwidth]{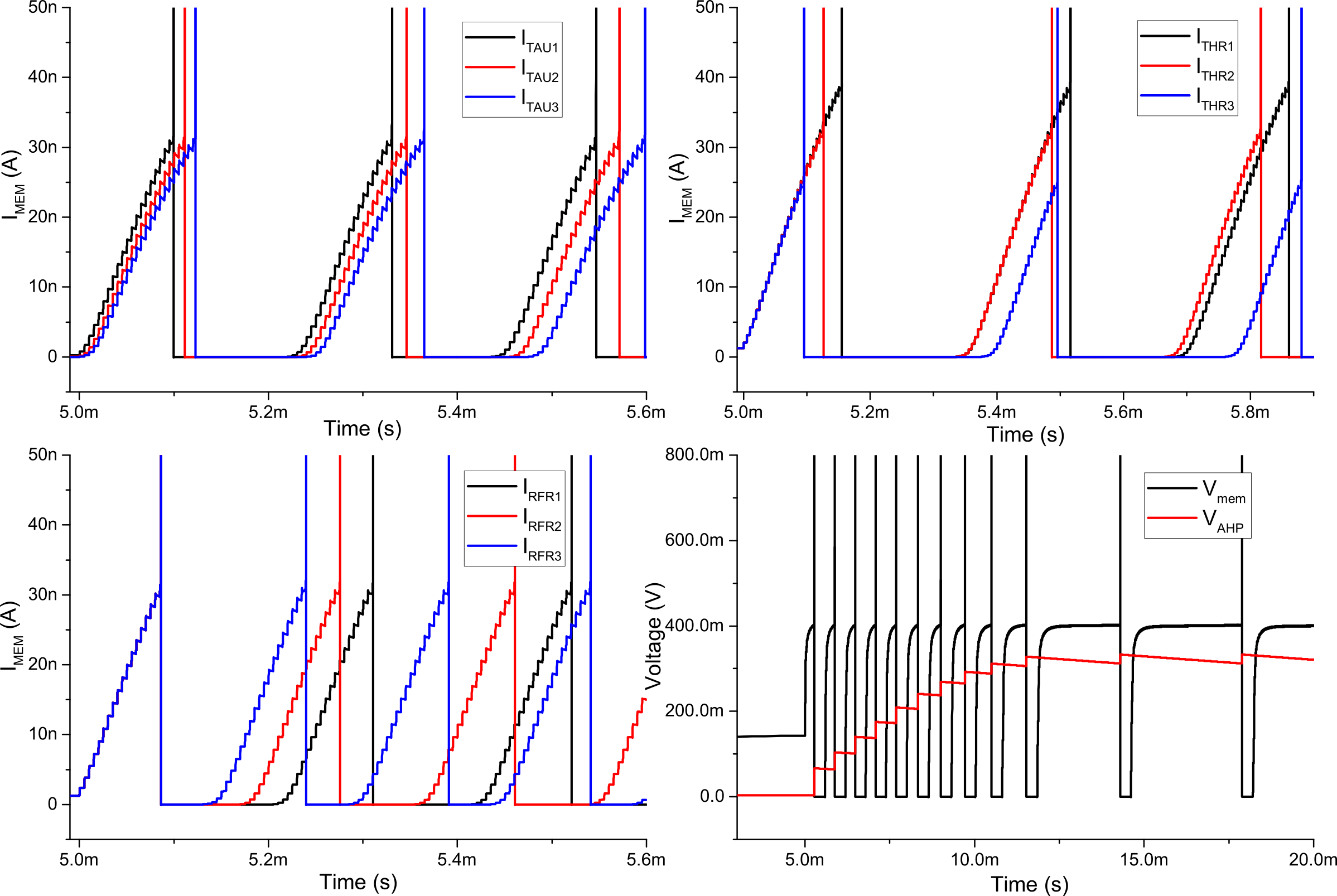}
  \caption{Different biologically plausible neuron's behaviors: (up-left) leaking time constants, (up-right) tunable firing threshold, (down-left) tunable refractory period duration, (down-right) spike-frequency adaptation. }
  \label{fig:neuron_sim}
\end{figure}

Simulation results demonstrating examples of biologically plausible behaviors are shown in Fig.\ref{fig:neuron_sim}. The top-left quadrant shows neuron membrane potential in response to a regular current spiking train for different leaking time constants ($I_{TAU3}>I_{TAU2}>I_{TAU1}$). The top-right quadrant shows the neuron response to a regular current spiking train injection for different values of firing threshold voltage ($I_{THR1}>I_{THR2}>I_{TTH3}$). The bottom-left quadrant shows the neuron response to a regular current spiking train injection for different settings of its refractory period ($I_{RFR3}>I_{RFR2}>I_{RFR1}$). The bottom-right quadrant demonstrates the spike-frequency adaptation behavior, obtained  by appropriately tuning the relevant parameters in the \textrm{AHP} block of Fig.~\ref{fig:soma} and stimulating the neuron with a constant injection current. 

With the transistor sizes chosen to minimize leakage current and device mismatch, the area of neuron, excluding the capacitor, is of approximately $20um^{2}$. The membrane capacitor can be overlayed onto the neuron layout using \ac{MIM} structures.  If we assume a capacitive density of approximately $18fF/um^{2}$, the area required to implement sufficiently large capacitors (e.g. $C_{M} = 0.5pF$, $C_{A} = 0.2pF$ and $C_{R} = 0.2pF$) will be approximately $50um^{2}$. 

\subsection{Asynchronous \acs{PCHB} digital circuits}

\ac{PCHB}-based asynchronous \ac{AER} routing/communication circuits are used to implement multi-core neuromorphic computing architectures. These circuits can be composed by combining  basic building blocks that implement basic processes (merge, split, buffer, etc.), and that follow a standard 4-phase hand-shaking protocol. 

\begin{figure}
  \centering
  \includegraphics[width=0.39\textwidth]{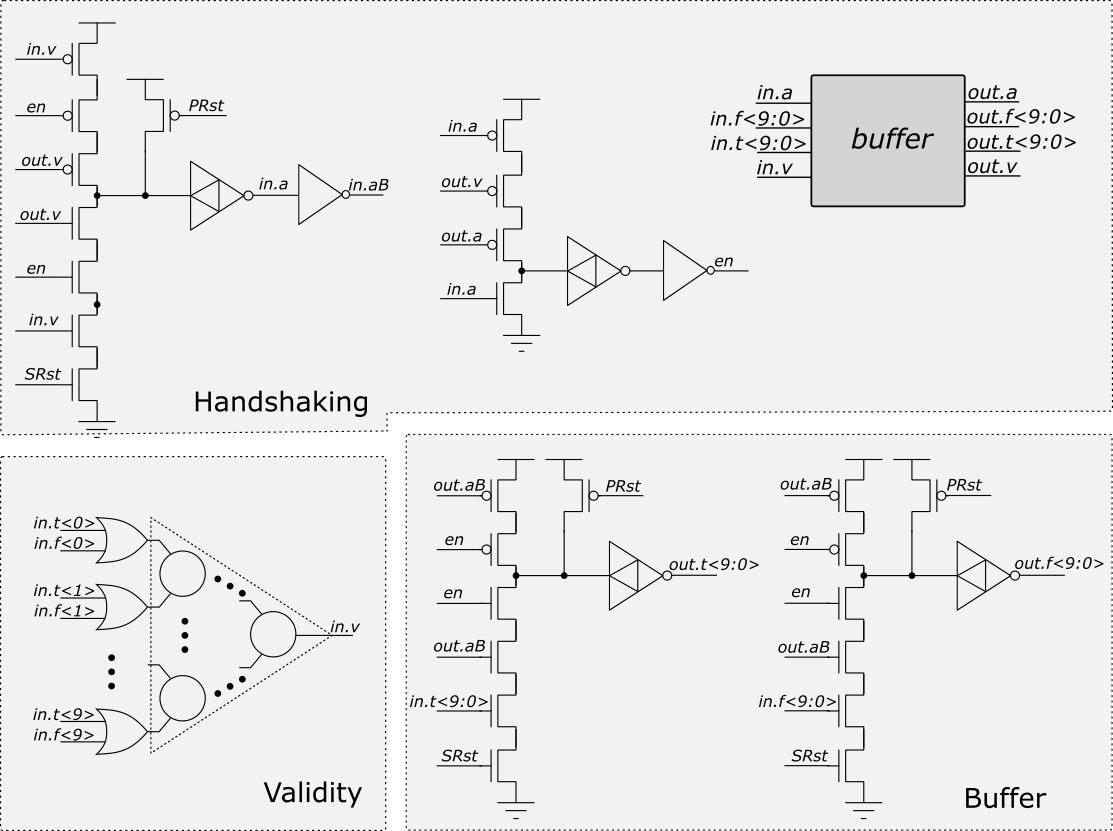}
  \caption{A \ac{PCHB} based \ac{QDI} buffer stage, which includes \emph{Handshaking}, \emph{Validity} and \emph{Buffer} blocks.} 
  \label{fig:buffer}
\end{figure}

Figure~\ref{fig:buffer} shows an example of RTL level pipeline buffer following 4-phase handshaking protocol and dual-rail protocol based on \ac{PCHB}. The process stage includes \emph{Handshaking}, \emph{Validity} and \emph{Buffer} blocks. With dual-rail data protocol, the request signal from a previous stage is encoded in data, the \emph{Validity} module checks the validity of input data and identifies the state via the signal $in.v$. The \emph{handshaking} block generates the acknowledge signal $in.a$ to acknowledge its previous stage for valid input. In parallel it will wait for the acknowledge signal from its following process stage, e.g. $out.a$, for its valid output. While taking care of hand-shaking with neighbour stages, the \emph{handshaking} block will generate control signal $en$ to enable \emph{Buffer} block for dealing with current input data or reset the \emph{Buffer} block for the next cycle (see Fig.~\ref{fig:buffer}).

Additional \ac{QDI} processes can be implemented with similar \emph{Handshaking}, \emph{Validity} and specific \emph{Function} block. By using a  dual-rail data flow and 4-phase handshaking, it is possible to build larger routing systems with more complex functions,  properly placing and combining these concurrent processes. 

\begin{figure}
  \centering
  \includegraphics[width=0.37\textwidth]{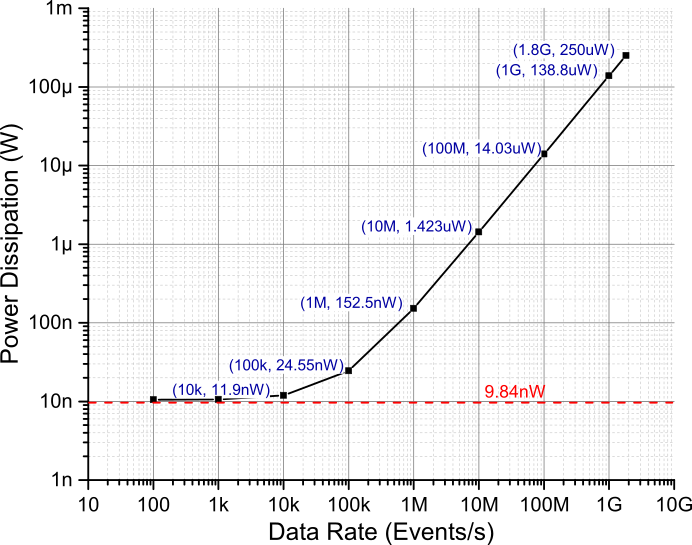}
  \caption{Power dissipation of a 10-bit PCHB-based buffer process versus input data rate.}
  \label{fig:power_buffer}
\end{figure}

Figure~\ref{fig:power_buffer} shows the power dissipation of a 10-bit PCHB-based buffer process versus different data rate of input events. The bandwidth of this 10-bit buffer is simulated to be 1.8\,G$\cdot$Events/s with a power dissipation of 250\,uW. The power dissipation of this buffer stage will scale down linearly corresponding to lower data rate. For a small data rate, e.g lower than 100k$\cdot$Events/s,  the static power dissipation will play a role. For a data rate smaller than 1\,k$\cdot$Events/s, the mean power dissipation will be dominated by static power dissipation, which is 9.84\,nW.  

In case of implementing large enough asynchronous \ac{AER} routing/communication system, according to estimates obtained from existing multi-core neuromorphic chips, the capacity of the whole asynchronous system will be equivalent of 600 10-bit buffers. In order to route 100\,k$\cdot$Events/s with this specific routing system, total power dissipation can be expected to be around 14.7\,uW with 147\,pJ per event. 

\subsection{CAMs for implementing configurable digital synapses}

\begin{figure}
  \centering
  \includegraphics[width=0.34\textwidth]{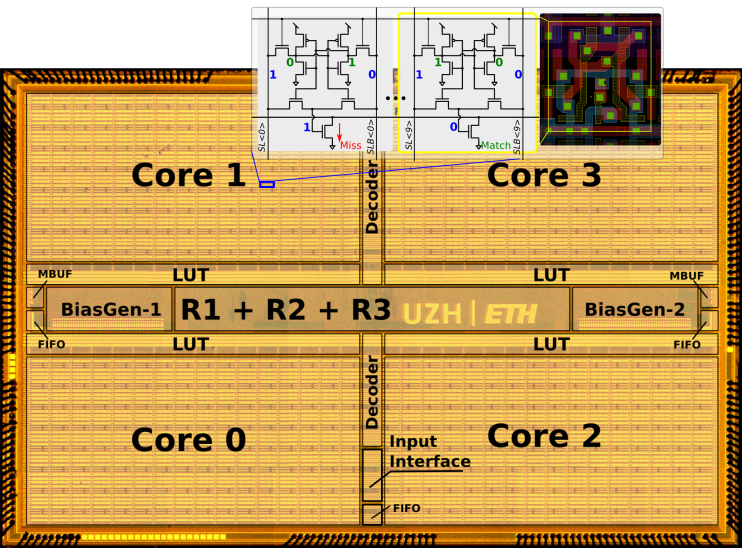}
  \caption{Multi-core neuromorphic processor fabricated in a 180\,nm CMOS process with an area of 43.79\,mm$^{2}$, comprising 1k neurons and 64k$\times$12-bit \ac{CAM} programmable synapses subdivided among 4 cores. Inserted figure shows CAM circuit and layout implementation.}
  \label{fig:CAM}
\end{figure}

Figure~\ref{fig:CAM} shows \ac{CAM} cells that have been used to implement configurable digital synapses in previous 180\,nm CMOS processes. The \ac{CAM} cell considered is based on NOR-type 9T cells and utilize a pre-charge-high Match-Line (ML) scheme. 

The layout of the CAM cell in 180\,nm bulk CMOS process occupies an area of $330F^{2}$. With advanced 28\,nm FPSOI process, it is reasonable to expect that the layout of \ac{CAM} will be more compact with a silicon area smaller than $330F^{2}$, which will result in less than $0.25um^{2}$. If we assume a fan-in of 64 programmable synapse connections that feed into one neuron (e.g., to implement convolutional networks with 8$\times$8 pixel kernels), and allocate 12-bits/synapse, the total silicon area that is  expected to be used is $192um^{2}$, which is much larger than the area of the analog neuron ($20um^{2}$). Therefore the silicon area of the multi-core neuromorphic processor in 28\,nm \ac{FD-SOI} process following architecture would be dominated by the \acp{CAM}.

In Table ~\ref{table:features} we compare features of multi-core neuromorphic processor with our old work in a 180\,nm CMOS process and a neuromorphic system in a 28\,nm CMOS process. 

\section{Conclusion}

In this paper we described some of the issues that have to be considered when scaling mixed-signal analog/digital neuromorphic circuits to advanced scaled process nodes. We showed how, by properly sizing the transistors of  analog neurons, and by optimizing its layout, it is possible to obtain reliable operation in a 28\,nm \ac{FD-SOI} process, reducing both silicon area and power consumption.
Furthermore, we showed that for asynchronous \ac{AER} routing system, scaling to more advanced processes leads to a significant improvement in bandwidth, but not to improvements in area and power efficiency. 
This however can be potentially solved by resorting to integration of \ac{RRAM} elements on the same substrate.




\begin{table}
  \centering
  \caption[]{Features of multi-core neuromorphic processor}
  \begin{tabular}{l | c | c | c}  
  \hline
               &  ~\cite{Indiveri_etal15}  & ~\cite{Mayr_etal16} & this work    \\ \hline  
    Technology & 180\,nm CMOS & 28\,nm CMOS & 28\,nm \ac{FD-SOI} \\ \hline 
    Supply voltage & 1.8V & 0.7V-1.0V & 1.0V \\ \hline 
    Energy per spike & 883pJ @ 30Hz & 2.3nJ-30nJ &50pJ @ 30Hz\\   \hline 
    Energy per routing & 360pJ & 230pJ & 147pJ \\  \hline 
      Bandwidth of routers & 400M$\cdot$Events/s  & 20M$\cdot$Events/s & 1.8G$\cdot$Events/s \\  \hline 
    Area of neuron & 1188$um^{2}$ & 64.6$um^{2}$ & 20$um^{2}$ \\   \hline         
    Area of synapse &128.4$um^{2}$ & 13$um^{2}$ & 3$um^{2}$\\ \hline  
  \end{tabular} 
  \label{table:features} 
\end{table}
\section*{Acknowledgment}

This work is supported by the EU ICT NeuRAM3 687299 grant. 



%



\bibliographystyle{IEEEtran}

\end{document}